\documentclass[useAMS,useusegraphicx, usenatbib]{mn2e}
\usepackage{amssymb}
\usepackage{epsfig}

\def\etal{et al.\ }

\def\gne #1#2{\ \vphantom{S}^{\raise-0.5pt\hbox{$\scriptstyle#1$}}_
{\raise0.5pt \hbox{$\scriptstyle#2$}}}

\title[Origin of Pseudobulge Hosting S0s]{Spiral Galaxies as Progenitors of
Pseudobulge Hosting S0s}

\author[Vaghmare \etal]
{Kaustubh Vaghmare$^{1}$\thanks{E-mail: kaustubh@iucaa.ernet.in (KV)}
Sudhanshu Barway,$^{2}$\thanks{E-mail: barway@saao.ac.za (SB)} 
Smita Mathur,$^{3,4}$\thanks{E-mail: smita@astronomy.ohio-state.edu (SM)}
\newauthor
and Ajit K. Kembhavi$^{1}$\thanks{E-mail: akk@iucaa.ernet.in (AKK)} \\
$^{1}$Inter University Centre for Astronomy and Astrophysics,  Post Bag 4,
Ganeshkhind, Pune 411007, India \\
$^{2}$South African Astronomical Observatory, P.O. Box 9, 7935, Observatory,
Cape Town, South Africa\\
$^{3}$Astronomy Department, The Ohio State University, Columbus, OH 43210, USA
\\
$^{4}$Center for Cosmology and Astro-Particle Physics, The Ohio State
University.}

\begin{document}

\maketitle

\label{firstpage}


\begin{abstract}
We present observations of pseudobulges in S0 and spiral galaxies using imaging
data taken with the Spitzer Infra-Red Array Camera. We have used 2-d
bulge-disk-bar decomposition to determine structural parameters of 185 S0
galaxies and 31 nearby spiral galaxies. Using the S\'{e}rsic index and the
position on the Kormendy diagram to classify their bulges as either classical or
pseudo, we find that 25 S0s (14\%) and 24 spirals (77\%) host pseudoblges.
The fraction of
pseudobulges we find in spiral galaxies is consistent with previous results
obtained with optical data and show that the evolution of a large fraction of
spirals is governed by secular processes rather than by major mergers. We find
that the bulge effective radius is correlated with the disk scale length for
pseudobulges of S0s and spirals, as expected for secular formation of bulges
from disk instabilities, though the disks in S0s are significantly smaller than
those in spirals. We show that early-type pseudobulge hosting spirals
can transform to pseudobulge hosting S0s by simple gas stripping. However,
simple gas stripping mechanism is not sufficient to transform the late-type
pseudobulge hosting spirals into pseudobulge hosting S0s.

\end{abstract}




\section{Introduction}

The Hubble tuning fork diagram \citep{Hubble1936} provides a snapshot of the
variety of morphologies found among galaxies at the present age. Despite
decades of research, a
number of questions remain unanswered concerning the processes responsible for
shaping them. In the last decade, several studies have especially focussed on
the central components of the late type galaxies. Originally, these central
components, known as bulges, were considered as elliptical galaxies surrounded
by a disk \citep{Renzini99} since they exhibited various signs of
massive elliptical galaxies, namely an older stellar population largely devoid
of dust, a dynamically hot, virialized system with a smooth structure and a
close to $R^{1/4}$ light profile. But subseqent observations showed that bulges
of many late-type galaxies exhibit almost opposite properties - younger
populations, rotational support, an exponential light profile and the presence
of structures generally found in disk viz. nuclear arms, spirals or bars. These
bulges are referred to as pseudobulges in the literature. (See
\citet{Kormendy2004} and references therein.)

These two types of bulges exhibit clear differences with respect to
various other
properties. For example, \citet{Drory2007} find that classical bulges are found
exlusively in the red sequence; \citet{Carollo2001} find a difference in the
mean optical-near-infrared $V-H$ color. Further, pseudobulges are known to obey
a well known correlation between the bulge effective radius and the disk scale
length \citep{Courteau96} while classical bulges do not or only weakly obey
such a correlation \citep{Fisher2008}. Pseudobulges, when viewed using high
resolution imaging instruments such as those available onboard Hubble Space
Telescope, exhibit structures such as spirals, bars or rings.
\citet{Fisher2008} use the presence of sub-structures as a working
definition of pseudobulges to show that most pseudobulges have a S\'{e}rsic
index $n<2$. Owing to these distinct characteristics of pseudobulges,
they are believed to have formed through secular processes within the
disk. This
interpretation is further supported by the fact that pseudobulges do not follow
several correlations otherwise obeyed by elliptical galaxies and classical
bulges, between absolute magnitude, S\'{e}rsic index, effective radius and
average surface brightness \citep{Gadotti2009, Fisher2008, Fisher2010}.
Pseudobulges may or may not have ongoing star formation in them
\citep{Fisher2010} but classical bulges in general do not show substantial
amount of star formation. All these studies also find that the transition of
properties from one bulge type to another is not sharp, possibly indicating the
existence of bulges with a mixture of properties.

Various mechanisms have been proposed to explain the formation of
pseudobulges.
These bulges are believed to have formed through secular processes
where presence of a structure breaking the axisymmetry of the disk drives gas
infall leading to the enhancement of the central stellar density which appears
as a bulge. \citep{Athanasoula1992, Noguchi99, Immeli2004,
Debattista2006, Elmegreen2009}. Studies such as \citet{Fisher2010} establish
that sufficient time may be available for secular evolution to grow bulges in
many cases, but there still exist pseudobulges where secular
processes alone cannot be responsible for observed bulge mass. To explain such
pseudobulges, cosmological simulations by \citet{Okamoto2013} and
\citet{Guedes2013} offer an alternate mechanism where the pseudobulges are
already present at earlier epochs as inner disks.  \citet{Keselman2012}
also showed that it is possible to form bulges that exhibit the photometric
signatures of pseudobulges and that such components can form through
major mergers of highly gas rich disk galaxies. 

Among the disk galaxies, pseudobulges occur more frequently in late type
spirals but recent studies \citep{Vaghmare2013, Weinzirl2009, Fisher2008} have
pointed out
that they can be found in a good fraction of S0s and early type spirals as well.
It is worthwhile therefore to study and search for any links between
S0 and spiral galaxies hosting pseudobulges. 

S0 galaxies are characterised by the presence of a central bulge and disk and
the absence of spiral arms. S0 galaxies occupy an important position on 
Hubble's tuning fork diagram \citep{Hubble1936}, where they are placed in
between
ellipticals and spirals implying that S0 galaxies have properties that are
intermediate to these two classes. Barway et al. (2007, 2009)
\nocite{Barway2007, Barway2009} have presented evidence to support the view
that the formation history of S0 galaxies can follow two very different routes.
Which route is taken appears to depend primarily upon the luminosity of the
galaxy, although the environment also plays a role. According to this
view, the
more luminous S0 galaxies have likely formed their
stars at an early epoch through hierarchical clustering wherein they undergo a
series of (minor) mergers or rapid collpase followed by star formation, as is
believed to be the case with elliptical galaxies \citep{Aguerri2005}. On the
other hand faint S0 galaxies have likely formed through secular processes. The
faint S0s would have likely originated from spiral galaxies which, in the
process
of their interaction with dense environments, had their star formation quenched
due to stripping of gas. These progenitor spirals could have formed their bulges
through secular processes such as the induction of gas inflows and vigorous star
formation by bars \citep{Salamanca2006, Bedregal2006, Barr2007}. In the context
of bulge dichotomy, this viewpoint suggests
that the more luminous S0 galaxies contain classical bulges while the fainter
ones contain pseudobulges \citep{Kormendy2004}. 
 \citet{Vaghmare2013} presented the first
systematic study of pseudobulges in S0 galaxies with emphasis on signatures of
their evolutionary processes on their progenitor disks. They found
that pseudobulges in S0 galaxies indeed occur preferentially in the fainter
luminosity class. They find that the discs of pseudobulge hosts possess on
average a
smaller scale length and have a fainter central surface brightness than their
counterparts occurring in classical bulge hosting S0 galaxies.

In this paper, we have carried out a mid-infrared study of a large sample of S0
galaxies and compared them with a sample of spiral galaxies to investigate the
possible origin of the pseudobulges in S0 galaxies and to search for
connections between the two morphological types. Throughout this paper, we use
the standard concordance cosmology with $\Omega_M= 0.3$, $\Omega_\Lambda= 0.7$
and $h_{100}=0.7.$


\section{Sample and Data Analysis}

\subsection{Description of the Samples}

The present study is based on two samples - (i) a sample of 185 S0 galaxies
used by \citet{Vaghmare2013} and (ii) a sample of 31 spiral galaxies
being studied by the authors as a part of an independent multiwavelength study.
These samples are described below.

The sample of S0 galaxies was constructed by \citet{Vaghmare2013} from a parent
sample comprising 3657
galaxies classified as S0 (having Hubble stage $T$ between -3 and 0) in the RC3
catalogue \citep{deVauc91}. Of these 3657 galaxies, only 1031, those with total
apparent $B$-band magnitude $B_T < 14.0$ were selected and cross-matched with
data in the Spitzer Heritage Archive (SHA)\footnote{The Spitzer Heritage Archive
is maintained by the Spitzer Science Centre and is a public interface to all
archival data taken using the three instruments on board the Spitzer Space
Telescope.( http://irsa.ipac.caltech.edu/applications/Spitzer/SHA)}. The
authors chose 
to use the 3.6 $\mu m$ imaging data taken using the Infrared Array Camera
(IRAC) on board the Spitzer Space Telescope because of three major virtues
namely - (i) that the light at this wavelength is relatively free from the
effects of extinction due to dust, (ii) it better takes into account the
contribution of the low mass stars which due to their presence in large numbers,
are a true representation of the stellar mass of a galaxy and (iii) the
observations are space-based and thus largely free from the problems that
generally plague ground based infrared observations.

We crossmatched the magnitude limited parent sample with the data in the
Spitzer Heritage Archive and found data for 247 galaxies. The data come from
different observing proposals submitted by different investigators over the
lifetime of the Spitzer mission. In most cases, the observations were targeted
while in some cases, serendipitious. Not all imaging data for these 247
galaxies were of the desired quality and this study uses results based on a
subset of 185 galaxies which has a median redshift of ∼0.005 with a
standard deviation of ~0.002. The highest redshift in our sample of S0s is 0.06.
The study carried out by \citet{Vaghmare2013} 
using this sample demonstrated that disks of S0s with pseudobulges
undergo transformation which the authors attribute to the
secular processes responsible for constructing the bulge. In this study, the
authors explore an alternate explanation to the lowered disk luminosity and
scale length found in the case of S0s with pseudobulges. In this view, the
pseudobulge hosting S0s could have been pseudobulge hosting spirals which
underwent gas stripping to transform into S0s. To compare one view with the
other, a comparison sample of spirals is needed.

As a part of an independent multiwavelength study of late-type galaxies,
we had
a sample of spiral galaxies which  was
constructed from the Nearby Galaxy Catalog \citep{Tully1988}. From that we created
quasi-volume-limited
sample by taking all galaxies within 20Mpc with following filters: (1) close to
face-on with inclination less than 35$^{\circ}$, to ensure minimum obscuration
by the disk of the target  galaxy, (2) Galactic latitude $|b| > 30$ to avoid
obscuration and contamination from our own Galaxy, and (3) no known starburst or
AGN activity. We also avoided E and S0 galaxies brighter than $M_B=-18.50$ as
well galaxies with types later than Sdm. The final sample consists of 56
galaxies; this sample was used in Ghosh et al. (2008a,b),  Ghosh (2009), and
Mathur et al. (2010) to look for nuclear AGNs with X-ray observations. Among
all these galaxies, $3.6\mu m$ imaging data was available for 35 galaxies all
comprising of spirals, both early and late-type. 

Among the galaxies which formed a part of this independent
multiwavelength study of AGNs, there were four S0/a galaxies ($T\sim0$) which
are already a part of the sample of 185 S0 galaxies. We have chosen to treat
these galaxies as S0s and thus they have been removed from the sample of spiral
galaxies. Thus, our total sample of spiral galaxies has 31 objects.

\begin{figure}
 \centering
 \includegraphics[width=8cm]{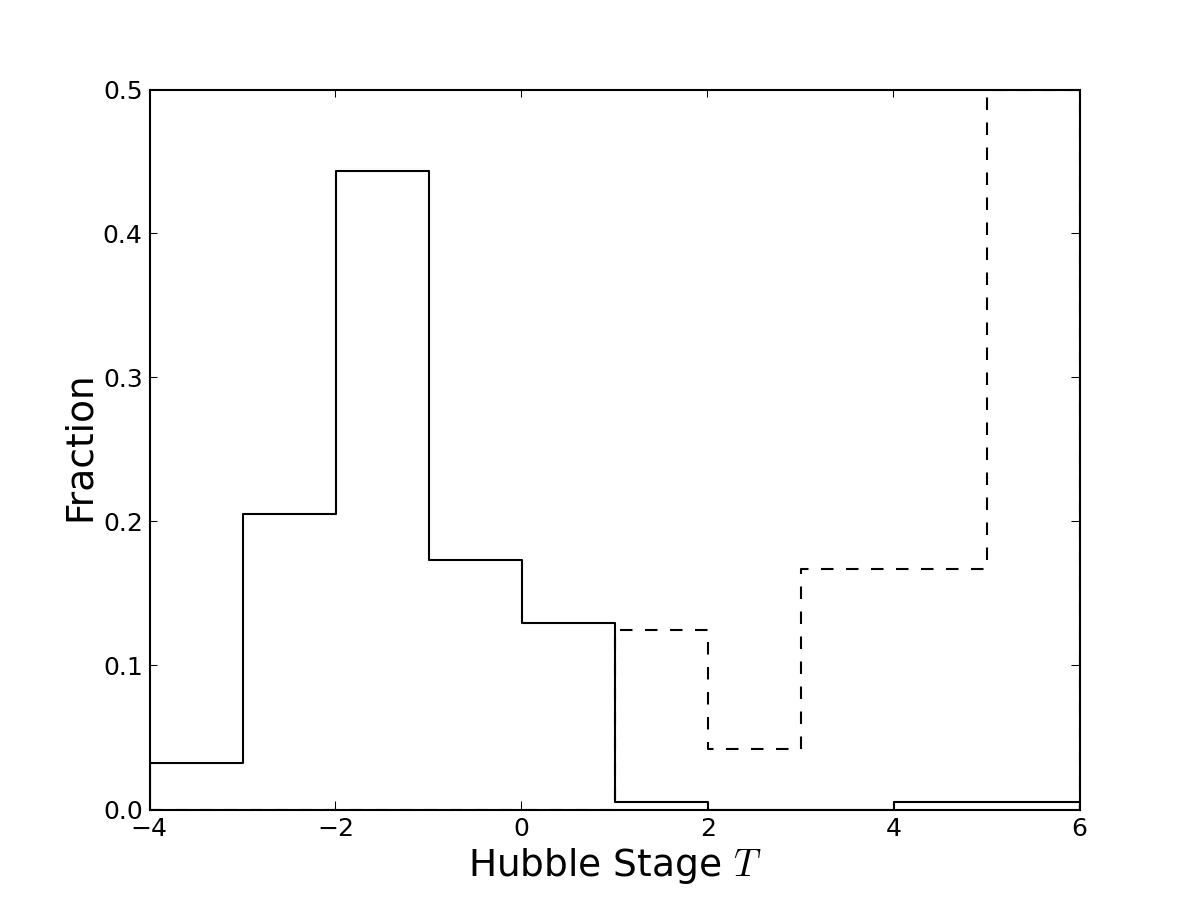}
 \caption{Distribution of the Hubble parameter $T$ for the sample of S0s
(solid line) and the spirals (dotted line). The Hubble parameters were
determined from the Hyperleda database and need not reflect the exact
morphology.}
 \label{fig:hubblestages}
\end{figure}

In Figure \ref{fig:hubblestages}, we plot the distribution of the Hubble stage
parameter $T$ for both the samples. As can be seen, our sample of 31 spiral
galaxies has both early and late-type spiral galaxies i.e. they reasonably
cover all the Hubble parameter range. The $T$ parameters values
used in this figure have been obtained 
from the Hyperleda database\footnote{http://leda.univ-lyon1.fr} and thus are
not necessarily a reflection of the exact morphologies. However, they are
sufficient for the purpose of a statistical comparison as in
Figure \ref{fig:hubblestages}. We have carried out
visual inspection of all the galaxies to ensure that our sample of S0s does not
contain any galaxies clearly containing spiral structure. The four galaxies
found common to parent samples do show a hint of faint spiral structure in
the deep Spitzer IRAC imaging data used in this study  but
we have chosen to treat them as transition objects and keep them in the sample
of S0 galaxies as they were a part of the study done by \citet{Vaghmare2013}.

\begin{figure}
 \centering
 \includegraphics[width=8cm]{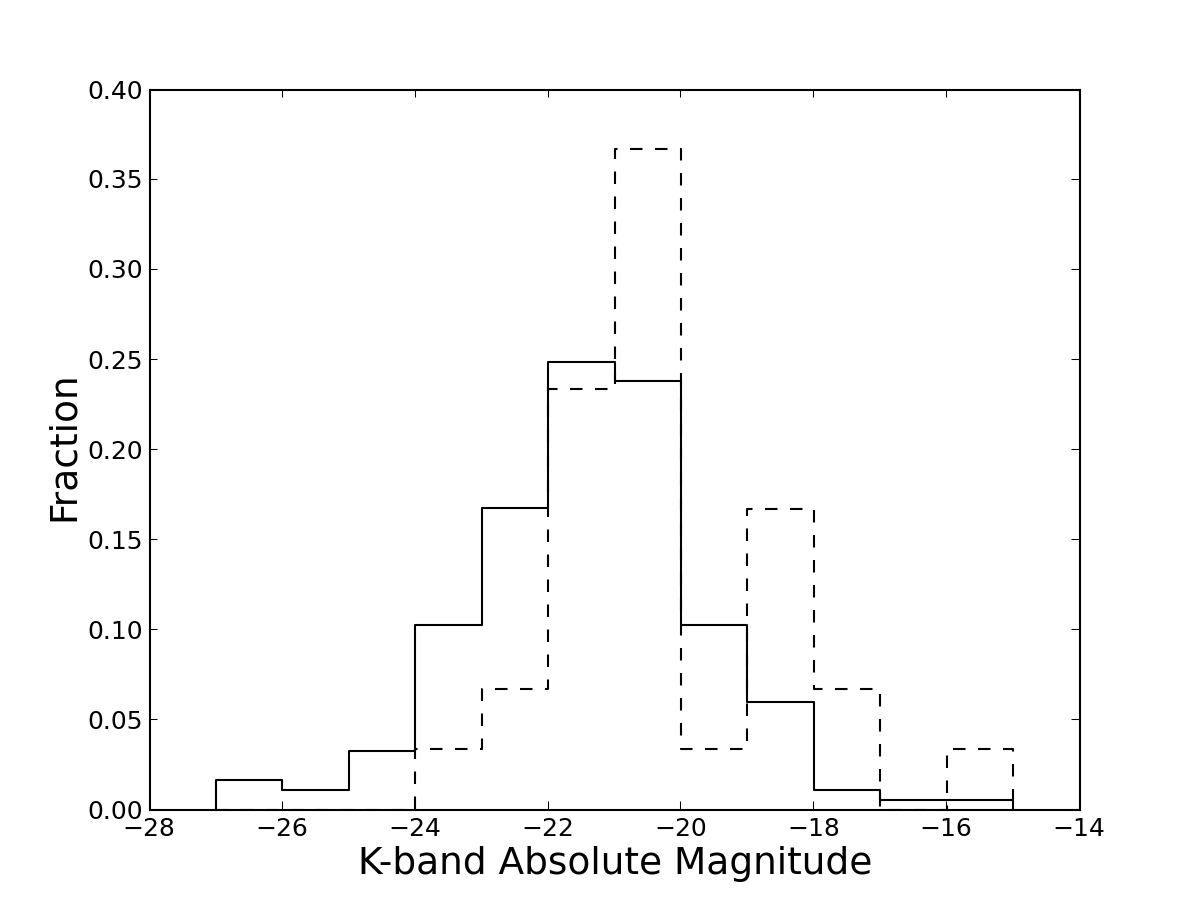}
 \caption{Distribution of the absolute K-band magnitude for the sample of S0s
(solid line) and the spirals (dotted line).}
 \label{fig:luminosity}
\end{figure}

While the sample selection criteria of S0s and the comparison sample of
spiral
galaxies are different, we found reasonable similarity in the final properties
of these samples which motivated us to compare the two.
Shown in Figure \ref{fig:luminosity} is the distribution of the
$K_s$-band absolute magnitude for both samples. The data used in plotting this
distribution were derived from the 2MASS Extended Source Catalog. The integrated
apparent magnitudes were converted into absolute magnitudes using redshift
information obtained from the NASA Extragalactic Database
(NED)\footnote{http://ned.ipac.caltech.edu/}. Note that the 2MASS catalog uses
the Vega system of magnitudes and thus these were converted to the AB system
\citep{OkeGunn83} by adding 1.84 to the Vega magnitude as suggested by
\citet{Munoz2009}. Unless
otherwise specified, we will use the AB magnitude system throughout the paper.
The distributions indicate that both samples have similar near-IR luminosities
and hence stellar mass. Further, as demonstrated in the subsequent sections,
the final number of pseudobulges in the two samples are comparable. These
properties motivate us to use the described sample of spiral galaxies as a
comparison sample. We note that while neither of these samples is complete, they
are representative of massive S0s and spirals in the nearby universe.

\subsection{Data Analysis Techniques}

\begin{table*}
\begin{center}
\begin{minipage}{166mm}
\caption{Best fit Bulge and Disk parameters for the pseudobulges in our S0
sample.}
\label{tab:params2}
\begin{tabular}{l c c c c c c c c c c c c c}
\hline
Name & $T$ & Bulge & $L_K$ & \multicolumn{3}{c}{Bulge parameters} &
\multicolumn{2}{c}{Disk parameters} & \multicolumn{3}{c}{Bar Parameters}  & B/T
& Bar/T \\
 & &  Type & & $<\mu_e>$ & $\rm{r}_e$ & $n$ & $\mu_{0d}$ & $\rm{r}_d$  &
$\rm{m}_{bar}$ & $n_{\rm{bar}}$ & $\rm{r}_{e(\rm{bar})}$ & & \\
\hline
\hline
ESO079-007 & -2.0 & PB & -18.73 & 23.25 & 0.87 & 0.53 & 21.23 & 1.72 & 0.00 &
0.00
& 0.00 & 0.04 & 0.00 \\ 
ESO085-030 & -0.4 & PB & -18.76 & 20.50 & 0.86 & 1.41 & 20.94 & 0.78 & 0.00 &
0.00 & 0.00 & 0.64 & 0.00 \\ 
ESO358-025 & -2.6 & PB & -19.04 & 20.42 & 0.73 & 0.74 & 21.72 & 1.59 & 0.00 &
0.00 & 0.00 & 0.41 & 0.00 \\ 
IC0051 & -2.0 & PB & -19.85 & 19.85 & 1.00 & 1.20 & 20.74 & 1.34 & 0.00 & 0.00 &
0.00 & 0.56 & 0.00 \\ 
IC2040 & -0.8 & PB & -18.70 & 20.58 & 0.98 & 0.83 & 22.26 & 1.44 & 0.00 & 0.00 &
0.00 & 0.68 & 0.00 \\ 
IC2085 & -1.2 & PB & -18.48 & 20.37 & 0.53 & 0.88 & 21.79 & 1.24 & 0.00 & 0.00 &
0.00 & 0.41 & 0.00 \\ 
NGC1510 & -1.8 & PB & -20.24 & 19.18 & 0.24 & 1.18 & 21.41 & 0.82 & 0.00 & 0.00
&
0.00 & 0.40 & 0.00 \\ 
NGC1522 & -2.3 & PB & -18.33 & 20.63 & 0.36 & 0.44 & 21.83 & 0.81 & 0.00 & 0.00
&
0.00 & 0.37 & 0.00 \\ 
NGC3413 & -2.0 & PB & -20.10 & 19.96 & 0.23 & 0.70 & 20.80 & 0.54 & 0.00 & 0.00
&
0.00 & 0.29 & 0.00 \\ 
NGC3773 & -1.7 & PB & -18.70 & 18.87 & 0.22 & 1.06 & 20.71 & 0.79 & 0.00 & 0.00
&
0.00 & 0.29 & 0.00 \\ 
NGC3870 & 0.6 & PB & -19.43 & 19.44 & 0.29 & 1.31 & 20.94 & 0.53 & 0.00 & 0.00 &
0.00 & 0.54 & 0.00 \\ 
NGC3990 & -2.1 & PB & -18.13 & 17.29 & 0.12 & 1.38 & 19.13 & 0.34 & 0.00 & 0.00
&
0.00 & 0.41 & 0.00 \\ 
NGC4336 & -0.6 & PB & -20.50 & 20.64 & 0.45 & 1.74 & 20.48 & 1.09 & 0.00 & 0.00
&
0.00 & 0.13 & 0.00 \\ 
NGC4460 & -2.0 & PB & -20.10 & 20.32 & 0.39 & 0.75 & 21.03 & 0.91 & 0.00 & 0.00
&
0.00 & 0.26 & 0.00 \\ 
NGC4544 & -2.0 & PB & -20.26 & 20.98 & 0.91 & 0.49 & 20.57 & 1.04 & 0.00 & 0.00
&
0.00 & 0.34 & 0.00 \\ 
NGC4880 & -1.9 & PB & -21.70 & 21.36 & 1.72 & 1.94 & 21.31 & 2.78 & 0.00 & 0.00
&
0.00 & 0.27 & 0.00 \\ 
NGC7371 & -1.6 & PB & -23.06 & 18.46 & 0.55 & 1.52 & 19.73 & 3.00 & 0.00 & 0.00
&
0.00 & 0.10 & 0.00 \\ 
NGC7709 & -1.9 & PB & -20.92 & 20.65 & 0.72 & 0.41 & 21.03 & 1.70 & 0.00 & 0.00
&
0.00 & 0.20 & 0.00 \\ 
IC0676 & -1.2 & PB & -20.02 & 20.62 & 1.14 & 1.13 & 20.89 & 1.71 & 14.07 & 0.05
&
0.29 & 0.30 & 0.16 \\ 
NGC1533 & 0.1 & PB & -18.79 & 16.73 & 0.22 & 0.95 & 19.77 & 1.58 & 12.94 & 0.50
&
0.89 & 0.22 & 0.07 \\ 
NGC3896 & -2.0 & PB & -21.33 & 21.54 & 0.53 & 0.66 & 22.77 & 1.65 & 16.65 & 0.01
& 0.34 & 0.23 & 0.05 \\ 
NGC4245 & -2.7 & PB & -20.34 & 17.32 & 0.24 & 1.23 & 20.28 & 1.58 & 12.86 & 0.35
& 1.66 & 0.22 & 0.14 \\ 
NGC4421 & -1.9 & PB & -21.15 & 17.35 & 0.24 & 1.74 & 20.59 & 3.03 & 13.12 & 0.64
& 1.77 & 0.10 & 0.15 \\ 
NGC4488 & -1.9 & PB & -19.45 & 19.59 & 0.36 & 1.36 & 24.09 & 5.48 & 12.58 & 0.75
& 2.50 & 0.12 & 0.43 \\ 
NGC5750 & -2.0 & PB & -21.62 & 17.84 & 0.32 & 0.98 & 19.94 & 2.56 & 13.64 & 0.20
& 2.23 & 0.09 & 0.09 \\ 
 \\
 \hline
\end{tabular}

\textbf{Notes:} Column 1 - The common name of a galaxy; Column 2 - Hubble
stage parameter $T$, Column 3 - Type of Bulge, CB=Classical Bulge and PB =
Pseudo-bulge,  Column 4 - K-band absolute magnitude (AB system), Column 5 - the
average surface brightness of the bulge within its effective radius, in mag per
arcsec$^2$, Column 6 - bulge effective radius in kpc, Column 7 - the bulge
S\'{e}rsic index, Column 8 - disk central brightness in mag per arcsec$^2$,
Column 9 - disk scale length in kpc, Column 10 - integrated apparent magnitude
of the bar, Column 11 - S\'{e}rsic index of bar, Column 12 - Bar effective
radius in kpc, Column 13 - Bulge-Total ratio, Column 14 - Bar-Total ratio.
\end{minipage}
\end{center}
\end{table*}

\begin{table*}
\begin{center}
\begin{minipage}{166mm}
\caption{Best fit Bulge and Disk parameters for our spiral sample.}
\label{tab:params}
\begin{tabular}{l c c c c c c c c c c c c c}
\hline
Name & $T$ & Bulge & $L_K$ & \multicolumn{3}{c}{Bulge parameters} &
\multicolumn{2}{c}{Disk parameters} & \multicolumn{3}{c}{Bar Parameters}  & B/T
& Bar/T \\
 & &  Type & & $<\mu_e>$ & $\rm{r}_e$ & $n$ & $\mu_{0d}$ & $\rm{r}_d$  &
$\rm{m}_{bar}$ & $n_{\rm{bar}}$ & $\rm{r}_{e(\rm{bar})}$ & & \\
\hline
\hline
IC5332 & 6.8 & PB & -21.30 & 21.54 & 0.62 & 1.39 & 21.48 & 2.62 & 0.00 & 0.00 &
0.00 & 0.05 & 0.00 \\ 
NGC1325A & 6.9 & PB & -17.74 & 20.98 & 0.24 & 0.97 & 21.25 & 1.86 & 0.00 & 0.00
& 0.00 & 0.02 & 0.00 \\ 
NGC3184 & 5.9 & CB & -22.41 & 19.49 & 0.33 & 2.93 & 20.73 & 3.17 & 0.00 & 0.00 &
0.00 & 0.03 & 0.00 \\ 
NGC3913 & 6.6 & CB & -19.88 & 21.70 & 1.19 & 2.46 & 22.89 & 2.20 & 0.00 & 0.00 &
0.00 & 0.47 & 0.00 \\ 
NGC3938 & 5.2 & PB & -22.51 & 18.97 & 0.32 & 1.11 & 19.76 & 1.84 & 0.00 & 0.00 &
0.00 & 0.06 & 0.00 \\ 
NGC4254 & 5.2 & CB & -25.77 & 20.72 & 10.06 & 3.44 & 19.89 & 6.27 & 0.00 & 0.00
& 0.00 & 0.55 & 0.00 \\ 
NGC4393 & 6.7 & PB & -16.44 & 23.60 & 2.66 & 1.39 & 24.63 & 7.67 & 0.00 & 0.00 &
0.00 & 0.24 & 0.00 \\ 
NGC4492 & 1.0 & CB & -22.91 & 19.09 & 0.81 & 4.45 & 20.43 & 2.63 & 0.00 & 0.00 &
0.00 & 0.25 & 0.00 \\ 
NGC4571 & 6.5 & PB & -19.93 & 20.72 & 0.23 & 1.37 & 20.69 & 0.98 & 0.00 & 0.00 &
0.00 & 0.05 & 0.00 \\ 
NGC4689 & 4.7 & PB & -23.86 & 19.67 & 0.66 & 1.13 & 20.34 & 4.06 & 0.00 & 0.00 &
0.00 & 0.05 & 0.00 \\ 
NGC5457 & 6.0 & PB & -22.17 & 19.03 & 0.29 & 1.86 & 20.31 & 1.70 & 0.00 & 0.00 &
0.00 & 0.08 & 0.00 \\ 
NGC628 & 5.2 & PB & -23.02 & 19.27 & 0.50 & 1.20 & 20.17 & 2.92 & 0.00 & 0.00 &
0.00 & 0.06 & 0.00 \\ 
IC5325 & 4.2 & PB & -23.30 & 20.63 & 3.46 & 0.73 & 20.45 & 1.11 & 15.24 & 1.38 &
0.33 & 0.88 & 0.02 \\ 
NGC1073 & 5.3 & PB & -22.21 & 21.59 & 0.57 & 0.69 & 21.81 & 5.61 & 13.52 & 1.10
& 2.37 & 0.01 & 0.07 \\ 
NGC1341 & 1.4 & PB & -22.31 & 21.45 & 1.94 & 0.82 & 19.76 & 1.23 & 13.79 & 0.50
& 0.94 & 0.27 & 0.22 \\ 
NGC1493 & 6.0 & PB & -22.09 & 19.88 & 0.11 & 1.18 & 20.72 & 2.31 & 14.78 & 0.86
& 1.08 & 0.00 & 0.04 \\ 
NGC1640 & 3.0 & PB & -22.91 & 17.95 & 0.36 & 1.50 & 20.41 & 2.49 & 13.05 & 0.44
& 2.45 & 0.14 & 0.19 \\ 
NGC1703 & 3.2 & PB & -22.87 & 18.59 & 0.29 & 0.84 & 20.47 & 2.86 & 14.34 & 0.05
& 2.30 & 0.05 & 0.05 \\ 
NGC255 & 4.1 & PB & -22.11 & 20.70 & 0.48 & 0.41 & 20.29 & 1.81 & 15.28 & 0.04 &
2.41 & 0.04 & 0.05 \\ 
NGC3344 & 4.0 & CB & -22.16 & 16.28 & 0.06 & 2.04 & 19.87 & 1.72 & 13.43 & 0.54
& 0.32 & 0.03 & 0.03 \\ 
NGC3887 & 3.9 & PB & -23.17 & 18.38 & 0.29 & 1.79 & 19.89 & 2.52 & 14.45 & 0.17
& 2.02 & 0.05 & 0.02 \\ 
NGC4136 & 5.2 & PB & -20.39 & 19.08 & 0.07 & 1.45 & 20.78 & 1.15 & 15.94 & 0.11
& 0.53 & 0.02 & 0.02 \\ 
NGC4314 & 1.0 & PB & -23.14 & 17.96 & 0.55 & 1.85 & 21.27 & 4.72 & 11.34 & 0.43
& 3.03 & 0.18 & 0.19 \\ 
NGC4394 & 3.0 & PB & -22.44 & 17.10 & 0.23 & 1.76 & 20.74 & 2.45 & 12.65 & 0.48
& 1.77 & 0.17 & 0.13 \\ 
NGC4411B & 6.3 & PB & -20.59 & 20.64 & 0.20 & 1.66 & 21.68 & 2.45 & 16.09 & 0.03
& 1.49 & 0.02 & 0.03 \\ 
NGC4688 & 6.0 & PB & -19.59 & 22.63 & 0.18 & 0.40 & 21.96 & 1.87 & 15.68 & 0.99
& 1.13 & 0.00 & 0.06 \\ 
NGC5068 & 6.0 & CB & -22.35 & 23.14 & 0.48 & 8.86 & 20.85 & 3.33 & 12.62 & 0.98
& 1.51 & 0.00 & 0.06 \\ 
NGC685 & 5.4 & PB & -22.27 & 21.55 & 0.70 & 0.86 & 21.33 & 4.49 & 15.06 & 0.04 &
1.87 & 0.02 & 0.02 \\ 
NGC7552 & 2.4 & CB & -24.28 & 15.10 & 0.31 & 0.34 & 19.91 & 2.98 & 11.46 & 0.31
& 3.84 & 0.38 & 0.22 \\ 
NGC991 & 5.0 & PB & -20.53 & 21.33 & 0.59 & 0.75 & 21.27 & 2.83 & 16.77 & 0.04 &
2.85 & 0.04 & 0.01 \\ 
UGC6930 & 6.6 & PB & -19.07 & 22.40 & 0.50 & 0.75 & 21.75 & 1.82 & 15.38 & 0.05
& 1.88 & 0.04 & 0.04 \\ 
\\
\hline
\end{tabular}

\textbf{Notes:} Column 1 - The common name of a galaxy; Column 2 - Hubble
stage parameter $T$, Column 3 - Type of Bulge, CB=Classical Bulge and PB =
Pseudo-bulge,  Column 4 - K-band absolute magnitude (AB system), Column 5 - the
average surface brightness of the bulge within its effective radius, in mag per
arcsec$^2$, Column 6 - bulge effective radius in kpc, Column 7 - the bulge
S\'{e}rsic index, Column 8 - disk central brightness in mag per arcsec$^2$,
Column 9 - disk scale length in kpc, Column 10 - integrated apparent magnitude
of the bar, Column 11 - S\'{e}rsic index of bar, Column 12 - Bar effective
radius in kpc, Column 13 - Bulge-Total ratio, Column 14 - Bar-Total ratio.
\end{minipage}
\end{center}
\end{table*}

The SHA serves three types of data products - Level 0 or raw, Level 1 also
known as Basic
Calibrated
Data (BCD) and Level 2 aka Post BCD. Typical observations of a galaxy using
Spitzer involve multiple dithered exposures which are later coadded to
form a mosaic. The individual dithered exposures are referred to as Level 1
data while the mosaic is referred to as Level 2 data. In our initial
analysis using Level 2 data, we found certain features which looked clearly
artificial, likely to be shortcomings in the way the mosaics were constructed as
opposed to being real features. We therefore chose to download the
Level 1 data and use the tool MOPEX (MOsaicking and Point source EXtraction)
provided by the Spitzer Science Centre to construct the Level 2 images. MOPEX
takes into account optical distortions, performs
image projection and outlier rejection to produce the final coadded image. We
found that the default settings of MOPEX were sufficient for our purposes.

In order to derive the structural parameters of the galaxies, we employed the
technique of two-dimensional decomposition of galaxy light using the program
GALFIT \citep{Peng2002}. The program GALFIT is versatile in terms of the
number and variety of components that can be fitted and uses the fast
Marquardt-Levenberg algorithm to minimise the $\chi^2$ and find the optimal
parameters for selected models. GALFIT requires an input image, a configuration
file containing a description of the model to be fitted, a mask image and the
information of the Point Spread Function (PSF). In order to construct the mask
images, we used the modified
segmentation images
created using SExtractor \citep{Bertin1996}. The PSF can be determined
using a variety of techniques. One can select bright field stars in the field
and co-add them or one can use utilities such as those available in the Image
Reduction and Analysis Facility (IRAF) to fit a suitable function to the PSF,
such as a Moffat profile. However, the Full Width at Half Maximum of the PSF in
case of IRAC images is between 1-2 pixels. Thus the use of such techniques is
not possible. We therefore used the instrumental PSF provided by the Spitzer
Science Centre. As the process of coadding the mosaics leads to a broadening of
the PSF, we convolved the instrumental PSF with a Gaussian whose parameters
were determined using neighboring bright field stars. The process of finding
the
 suitable Gaussian was done separately for each image.

An initial fit for the surface brightness propfiles of all galaxies was obtained 
by fitting for their bulge and disk components of the galaxies. The bulge was 
modelled using the S\'{e}rsic profile
\citep{Sersic68} given by the following equation,

\begin{eqnarray}
\centering
I_{bulge}(x,y)  &=& I_b(0) e^{  -2.303 b_n  (r_{bulge}/r_e)^{1/n}}, \\
r_{bulge} &=& \sqrt{x^2 + y^2/(1 - e_b)^2}, \nonumber
\end{eqnarray}

where $x$  and $y$  are the  distances from the  centre of  the galaxy
along the major and minor axis respectively while $e_b$ represents the
ellipticity of the bulge. The parameter $b_n$ is a function of
$n$, which can be written as $1.9992 n - 0.3271$ \citep{Capaccioli89}.
$n$ here, is the S\'{e}rsic index of the bulge which is also an 
indicator of the concentration of light, $r_e$ is the effective radius
of the bulge which is also known as the half-light radius as it contains
half the total integrated light in the bulge. $I_b(0)$ is the central intesnity
of the bulge.

The disk
was modelled using an exponential function which has scale length ($r_d$) and
central intensity ($I_{\rm{d}}(0)$) as the main free parameters.

\begin{center}
\begin{equation}
\centering
 I_{exp}(r_{\rm{disk}}) = I_{\rm{d}}(0) \exp \left( \frac{r_{\rm{disk}}}{r_d
}\right)
\end{equation}
\end{center}

All the parameters including
galaxy centre and sky background were left free. In particular, the sky
background was independently measured and in a few cases where the best-fit
value did not agree with the independent measurement, the fit was rerun with the
sky background held fixed to the determined value. In cases where the
residual image obtained by subtracting the PSF convolved model image from 
the observed image revealed a bar, a second run of fitting was performed with
an additional S\'{e}rsic included to account for the bar. The inclusion of a
bar allows the bulge parameters to be free of the systematics introduced by
an unaccounted bar as demonstrated in studies by \citet{Gadotti2008} and
\citet{Laurikainen2005}. The S\'{e}rsic profile describing the bar can be
identified
as that which has a relatively low axis ratio ($q \sim 0.1$) as well as
S\'{e}rsic index $n$. ($n \lesssim 0.5$).

A key limitation of the 2-d decomposition technique is that it cannot be
meaningfully employed in situations where the morphologies of the galaxies are
disturbed as a result of an ongoing merger or a strong tidal encounter.
Further, the
specific algorithm employed by GALFIT requires a reasonable signal-to-noise
ratio (SNR). Thus galaxies whose images had poor SNR, those with
disturbed
morphology and a few other galaxies for which meaningful solutions could not
be found were removed from the final sample. We have a final sample of 185 S0s
of which 63 (34\%) are barred and 31 spiral galaxies of which 19 (61.2\%) are
barred. We have not tried modelling the spiral arms as our primary
interest is in determination of the bulge and disk parameters.

A note on the reliability of decomposition for galaxies with small
angular sizes: \citet{Gadotti2008} have shown that the bulge properties can be
recovered reliability if the bulge effective radius is not lesser than 80\% of
the Half Width at Half Maximum (HWHM) of the PSF. In the case of S0 galaxies,
$\sim 15$ galaxies have a bulge with effective radius between 80\% and 100\%
the HWHM of the PSF while no bulges fall below the 80\% mark. For these 15
galaxies, we have carefully inspected the decomposition results and have
ensured that parameters derived are reasonable. Further, none of these galaxies
are a part of the pseudobulge subset and thus one can safely ignore possible
biases being introduced in comparison of pseudobulge properties, due to seeing
effects.

In the present paper, our primary focus is on the pseudobulges found in the S0s
and spiral galaxies. A detailed discussion on classical bulges in S0 galaxies
will be presented in Vaghmare \etal (2015, in preparation). The structural
parameters of the pseudobulge hosts in S0 galaxies are
provided in Table \ref{tab:params2} while those of the 31 spiral galaxies are
presented in the Table \ref{tab:params}.

\section{Classification of Bulges}

\subsection{A Brief Review of Classification Methods}

Pseudobulges can be identified based on the presence of substructures
typically found in disks, such as nuclear spiral arms, rings or bars
\citep{Carollo98}. \citet{Fisher2008} used high resolution imaging data from
the Hubble Space Telescope and classified bulges into classical and pseudo,
based on the presence of nuclear structure. They found that the bulges exhibit
a bimodal distribution of S\'{e}rsic \citep{Sersic68} indices $n$ with most 
bulges having $n<2$ being pseudobulges and those having $n>2$ being classical.
Several studies thus have used the S\'{e}rsic index to classify bulges
\citep{Okamoto2013, Fisher2010, Vaghmare2013}. 

Various other criteria can be used to distinguish between classical and
pseudobulges (see \citet{Kormendy2004} for a detailed review). Some of these
criteria are (a) a boxy shaped bulge (detectable in the case of an
edge-on galaxy), (b) kinematics dominated by rotation (determined by measuring
the ratio of rotational velocity and velocity dispersion i.e.
$V/\sigma$), (c)
deviation from the Faber-Jackson relation towards lower velocity dispersion and
(d) a dominant presence of Population 1 stars (but without any sign of a recent
merger). 

Any criterion leads to practical difficulties when applied to a broad sample.
Kinematic information can be obtained only using spectroscopic data and
collection of these for large number of objects is expensive. Another method is
to detect nuclear structure, but this requires high resolution images taken at
very low seeing, as provided by Hubble Space Telescope \citep{Fisher2008}. But
\citet{Fisher2008} found, after classifying bulges using such a method that a
division line of $n=2$ can be used to classify bulges with those $n<2$ being
pseudobulges. However, a key problem in using the S\'{e}rsic index alone to
classify bulges is that error bars on it can be quite large \citep{Vaghmare2013,
Gadotti2008}. 

An alternate classification scheme proposed by \citet{Gadotti2009} which tries
to address the problem associated with using S\'{e}rsic index alone is to
classify bulges based on their position on the Kormendy diagram
\citep{Kormendy77} which is a plot between the logarithm of the
effective radius of the bulge and its average surface brightness within the
effective radius. It
is one of the projections of the fundamental plane
\citep{Djorgovski87,Dressler87} of elliptical galaxies and also bulges of early
type disk galaxies \citep{Ravikumar2006}. Bulges evolving from disks are not
expected to obey the tight correlation found in case of elliptical galaxies.
\citet{Gadotti2009} classifies those bulges lying more than $3\sigma$ below the
best-fit line to the elliptical galaxies on this diagram as pseudobulges, where
$\sigma$ is the root-mean-square scatter in the correlation for ellipticals.
\citet{Gadotti2009} does not use the S\'{e}rsic index in classifying bulges but
demonstrates that it is possible for pseudobulges to exist, as classified by
them, with $n>2$ \citep{Gadotti2009, Mathur2012,Xivry2011}.

\subsection{Identifying Bulges in Our Samples}

In various studies carried out in literature, any bulge that has evolved
through secular
processes is regarded as a pseudobulge. As shown by various studies highlighted
in the previous subsection, this is equivalent to calling pseudobulges as those
that satisfy various observational crtieria discussed in the prior
section.
More the criteria used for bulge classifcation, the more secure the result will
be, so we chose to classify bulges as pseudobulges if they had $n<2$ and were
also more than $3\sigma$ below the best-fit line to ellipticals on the Kormendy
diagram. This classification scheme selects fewer pseudobulges but as argued in
\citet{Vaghmare2013}, is more secure. (Refer to Table 3 for
a comparison of pseudobulge statistics obtained using different criteria.)
It was not possible to use the presence of nuclear structure or kinematics
for bulge classification as the required data for the same were not
uniformly available for our samples. 

\begin{figure}
 \centering
 \includegraphics[width=8cm]{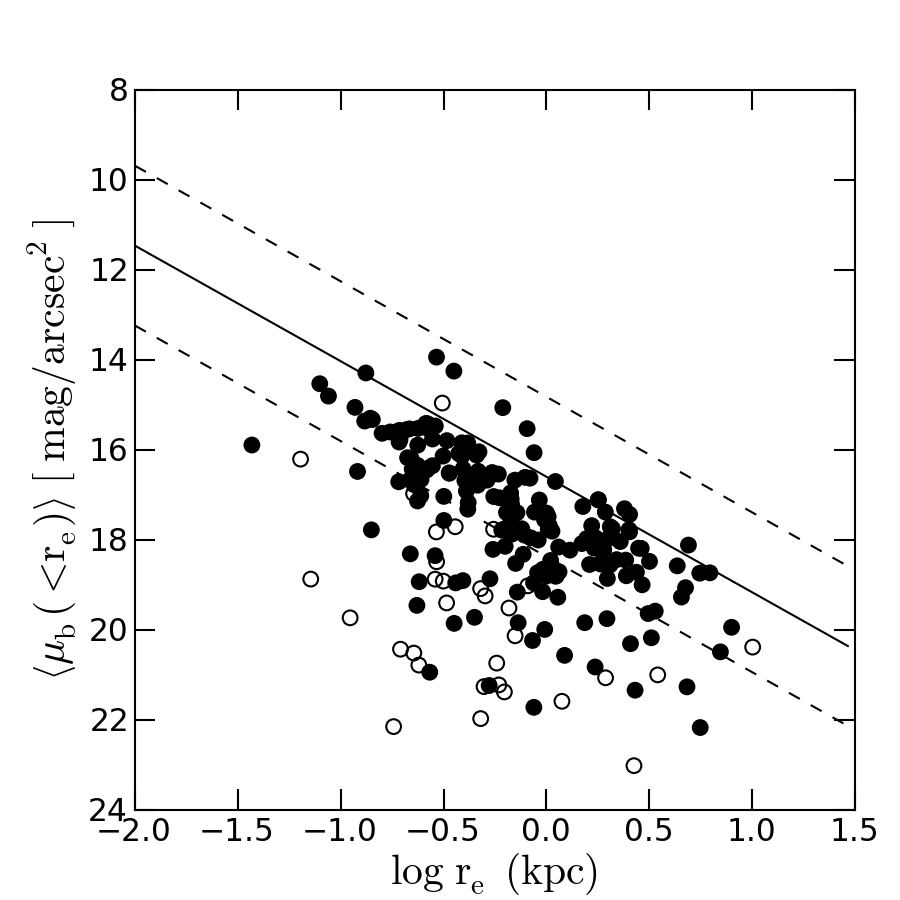}
 \caption{The Kormendy diagram showing the position of the bulges of S0
galaxies (filled circles) relative to the bulges of our sample of spiral
galaxies (empty circles). The solid line is the
best-fit to ellipticals  while the dashed lines mark the $3\sigma$
boundaries. Points representing elliptical galaxies have been left out for
clarity.}
 \label{fig:kormendy}
\end{figure}

To determine the best-fit line to the ellipticals, we used the decomposition
data from a study by \citet{Khosroshahi2000} of Coma cluster ellipticals done
in the K-band. The magnitudes reported in this study were first transformed to
the 3.6 $\mu m$ band of Spitzer IRAC using the relation $K - m_{3.6} = 0.1$
\citep{Toloba2012, Falcon2011} and then to the AB magnitude system by adding
1.84 following \citet{Munoz2009}. The best-fit line is shown in Figure
\ref{fig:kormendy} but for clarity, the points representing elliptical galaxies
have been left out. Also shown on this plot are the bulges of S0 galaxies
(filled circles) and those of spiral galaxies (empty circles). One notices an
offset between bulges of S0s and the best-fit line to the ellipticals, which is
consistent with the findings of \citet{Barway2009}. The bulges of spiral
galaxies exhibit a greater offset, which is consistent with studies such as
\citet{Ravikumar2006, Laurikainen2010}. 

Among 185 S0 galaxies, 25 are classified as pseudobulge hosts using the
above criterion. For the same data set, in \citet{Vaghmare2013} we
had reported 27 galaxies as pseudobulge hosts. During the course of further
study, we updated the structural parameters and found that two of these were no
longer
classified as pseudobulges. This change in number of pseudobulges
does not affect the essential findings of \citet{Vaghmare2013}. In case of
spiral galaxies, we find 24 of 31 to be pseudobulges. In Table 3, we summarize
the distribution of the bulge types in the two samples as found using various
criteria, for comparison. For the remainder of the study, we use the bulge
classification based on both S\'{e}rsic index and the Kormendy relation to
identify bulge type.

\vspace{0.5cm}

\begin{table}
\caption{Distribution of the bulge types in the two samples according to various
criteria.} 
\begin{center}

\begin{minipage}{65mm}

\label{tab:bulgeclass}
\begin{tabular}{llrr}
\hline
Criterion Used    &   Bulges     & S0's    & Spirals \\
\hline 
Kormendy only   &  Classical    & 137   & 2    \\
                           &  Pseudo      &   48   & 29  \\
\hline
S\'{e}rsic only     &  Classical    & 111   & 6     \\
                           &  Pseudo      &   74   & 29   \\
\hline
Both                    &  Classical    & 160  & 7     \\
                           &  Pseudo      &   25   & 24   \\

\hline 
\hline
\end{tabular}
\end{minipage}
\end{center}
\end{table}


\section{Results}

\begin{figure}
 \centering
 \includegraphics[width=8cm]{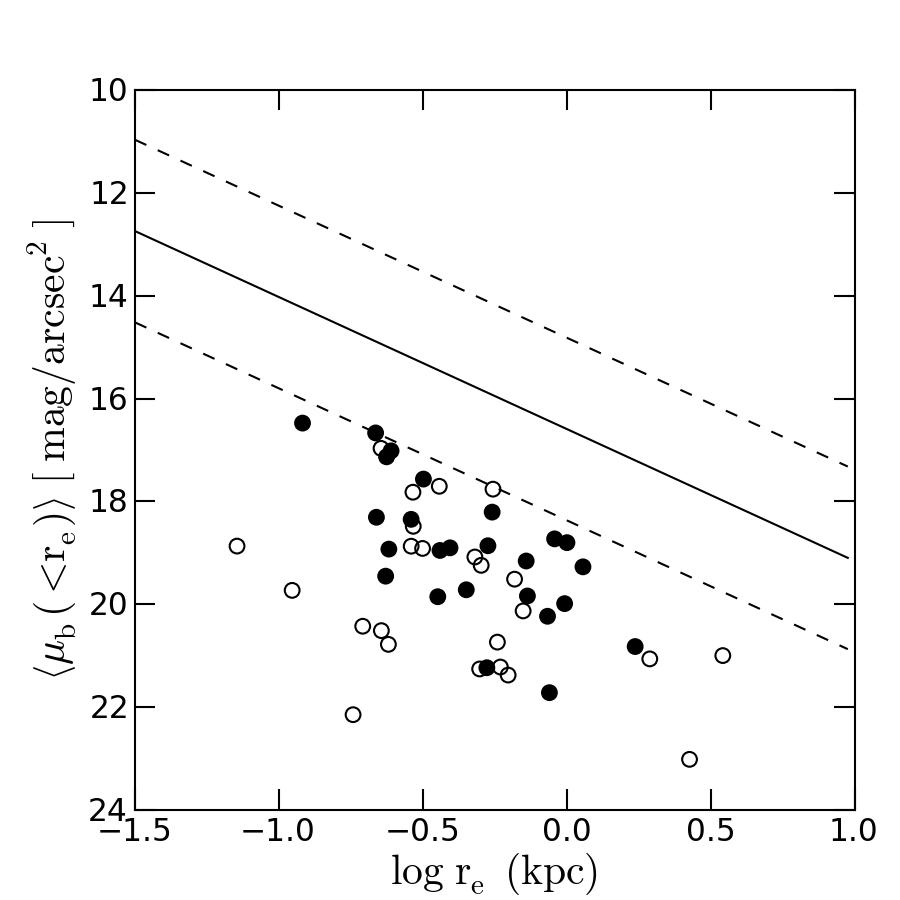}
 \caption{Same as Fig. \ref{fig:kormendy}, but plotting only the psuedobulges
identified using both the criteria discussed in the text, for S0s (filled
circles) as well as spirals (open circles).}
 \label{fig:kormendy_pbs}
\end{figure}

Our primary motivation in this study is to try and systematically compare the
pseudobulges in S0s with those in spirals and find clues to the origin of
pseudobulges.
In Figure \ref{fig:kormendy_pbs}, we replot the Kormendy diagram showing only
the pseudobulges of both samples. This plot reveals an interesting feature.
Along the effective radius axis, one does not find any appreciable difference
in distribution
between the pseudobulges of the two morphological classes. Along the average
brightness axis though, one finds that pseudobulges of spiral galaxies tend to
be fainter on average. The difference in the mean surface brightness is $\sim
0.9$ $\rm{mag/arcsec^2}$ with a significance of more than 95\% as determined
using a \emph{t-test}. 

An often discussed correlation in the context of the secular evolution scenario
of the galaxy is that between the bulge effective radius $\rm{r}_e$ and the
disk scale length $\rm{r}_d$ \citep{Courteau96}. \citet{Khosroshahi2000}
have also studied this correlation for different types of disk galaxies.
Models or simulations where the
bulge forms via a major merger and the disk forms later via gas accretion do not
predict a strong correlation between these parameters. However, such a
correlation is expected in a scenario where the disks form first and internal
processes then build up the bulge by rearranging disk material.
\citet{Barway2007}
found that bright S0s show a weak anticorrelation while the fainter S0s show a
strong positive correlation between $r_e$ and $r_d$. The authors
concluded that fainter S0s likely evolve through secular processes. 
\citet{Vaghmare2013} also plotted $r_e$ against $r_d$ for all S0s and
found that pseudobulge hosts possess a lower scale length than their
counterparts hosting classical bulges.

\begin{figure}
 \centering
 \includegraphics[width=8cm]{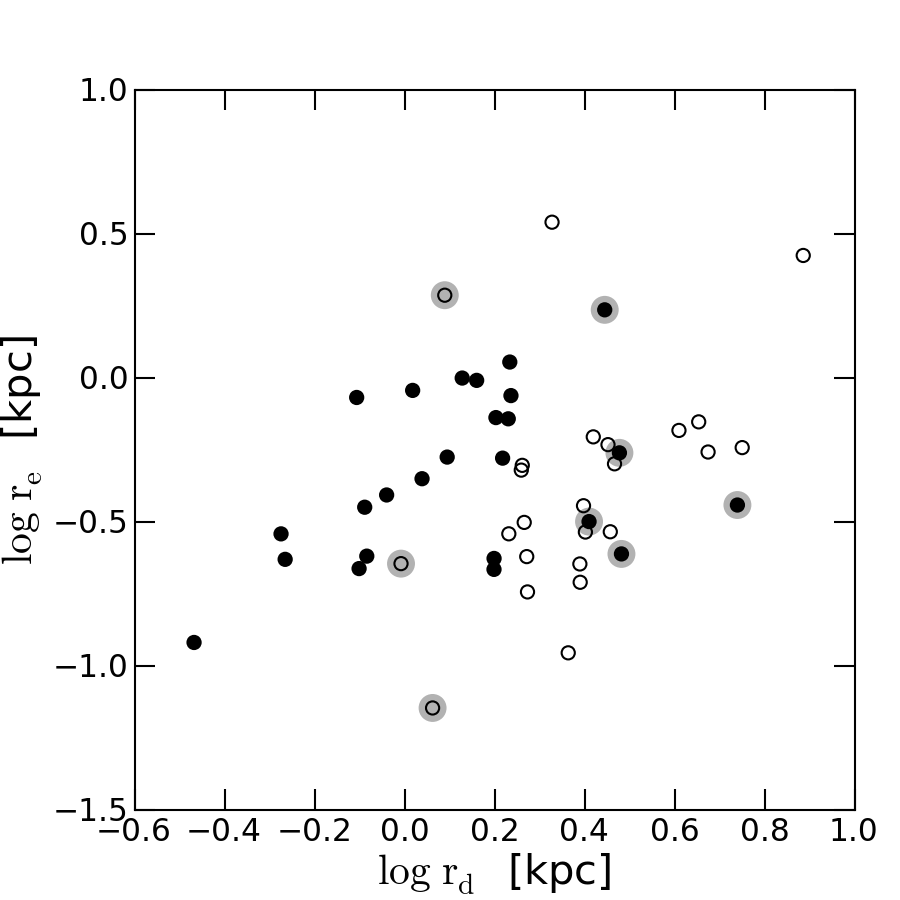}
 \caption{A plot of bulge effective radius against the disk scale length for
the pseudobulges in our sample of S0 galaxies (filled circles) and spiral
galaxies (empty circles). The outliers in both the samples are higlighted in
grey.}
 \label{fig:rerd_pbs}
\end{figure}

In Figure \ref{fig:rerd_pbs}, we have plotted this diagram for the
pseudobulges of both our samples. The pseudobulges in both the
samples show a correlation albeit with an offset. For a given $r_e$, one sees
that on average the disk scale length $r_d$ in the case of S0s is smaller than
in
case of the spiral galaxies. The mean disk scale length for S0 galaxies is 1.6
kpc while for spirals it is 2.8 kpc. The
significance level of this difference is more than 99.9\%.  

\begin{figure}

 \centering

 \includegraphics[width=8cm]{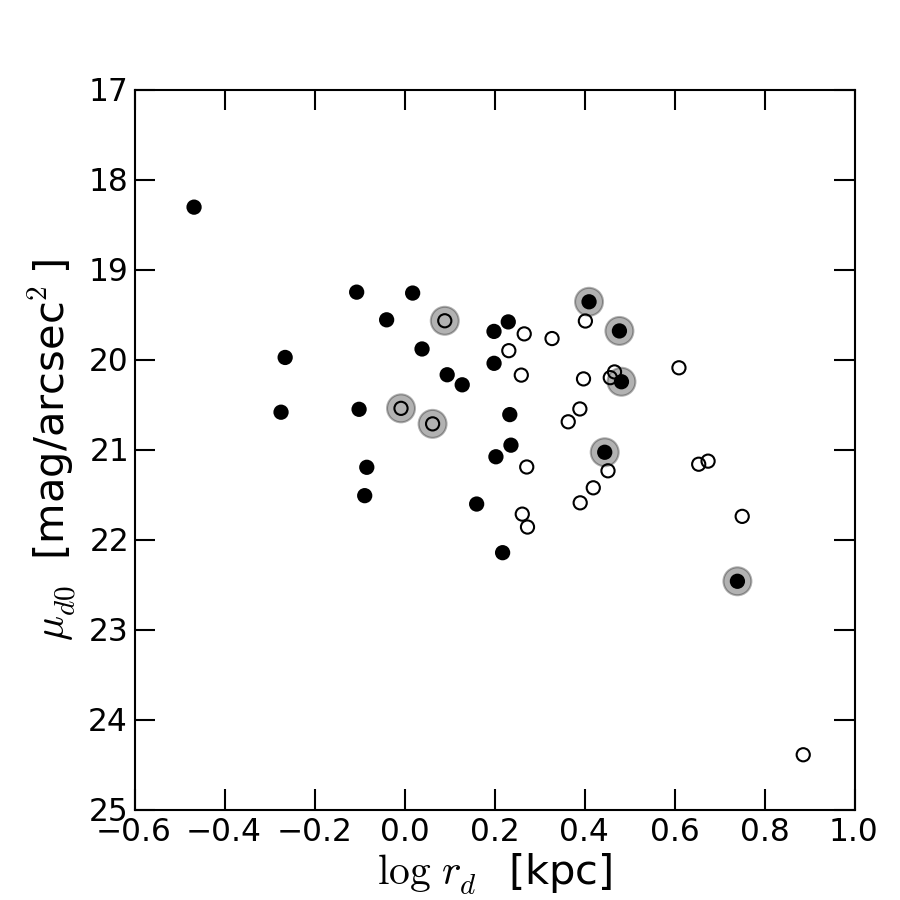}
   \caption{A plot of disk central surface brightness against the disk scale
length for
the pseudobulges in our sample of S0 galaxies (filled circles) and spiral
galaxies (empty circles). The outliers described in Section 4 are shaded in
grey.}
 \label{fig:mu0d_rd}

\end{figure}

\begin{figure*}
 \centering
 \includegraphics[width=14cm]{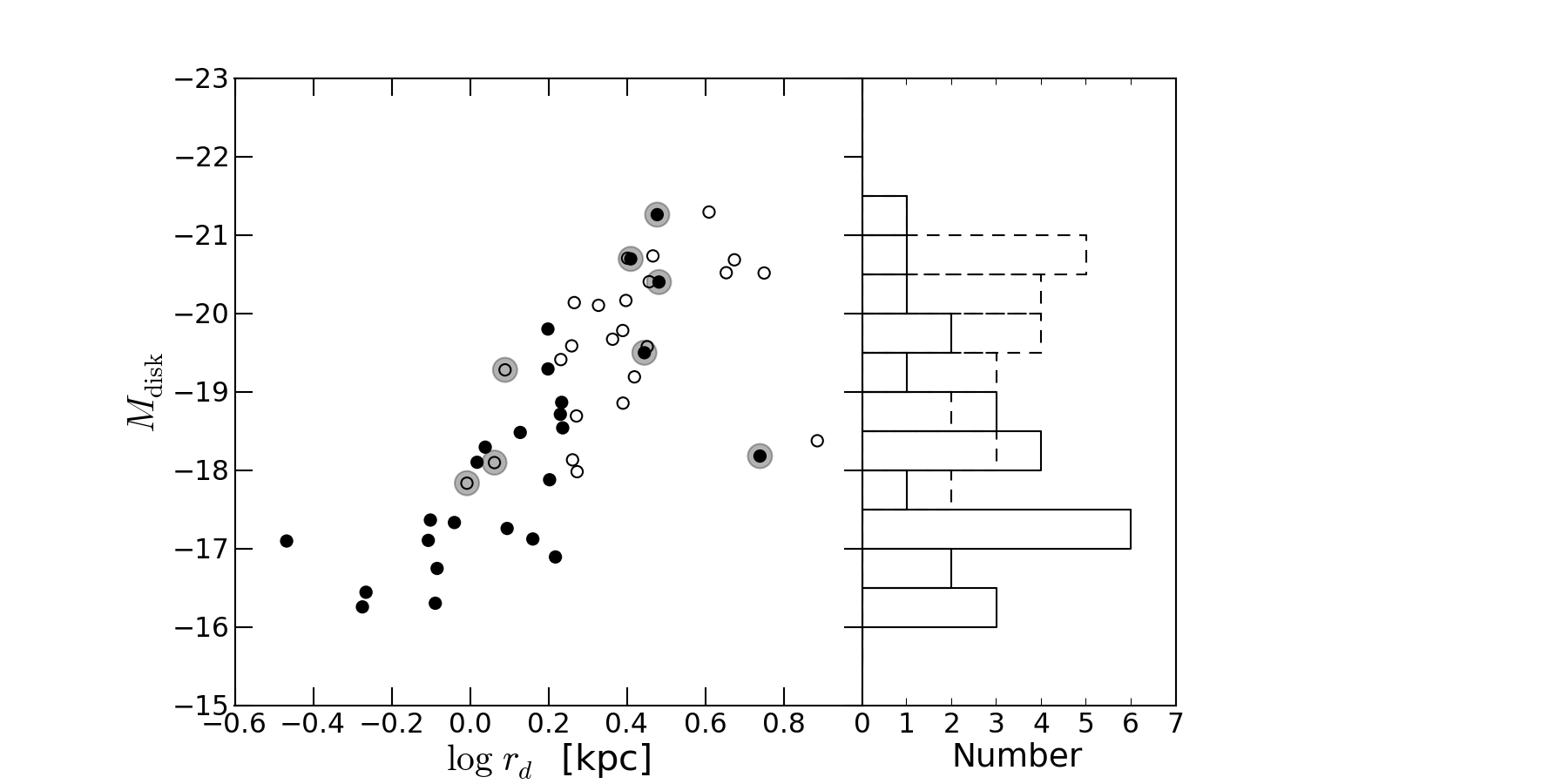}
   \caption{A plot of disk absolute magnitude against the disk scale
length for the pseudobulges in our sample of S0 galaxies (filled circles) and
spiral galaxies (empty circles). Also shown in the right panel is the
distribution of disk absolute magnitude with solid lines representing S0s while
dotted lines representing spirals. The outliers described in Section 4 are
shaded in grey in the left panel.}
 \label{fig:dabsmag_rd}
\end{figure*}

It is seen that there are outliers on either side i.e. there are S0s
which possess a scale length typical of spirals and vice versa. We now comment
on these outliers, which for convenience, have been shaded in
grey in Figure \ref{fig:rerd_pbs}. \textit{The same outliers have also been
marked in Figures \ref{fig:mu0d_rd} and \ref{fig:dabsmag_rd}.} The five S0
galaxies found
``amidst''
the
spiral galaxies on the
$r_e - r_d$ diagram are NGC 4488, NGC 4421, NGC
7371, NGC4880 and NGC 5750. Of these NGC 4488 and NGC 4421 are classified
as S0/a
galaxies with $T\sim0$ and are known to reside in the Virgo cluster. NGC 7371
and NGC 5750 are also classified as S0/a galaxies, with NGC 5750 known to be a
part of a rich group environment while reliable environment information for NGC
7371 is not available. Together, these four objects are the same objects in
our sample which were initially found to be common to the sample of S0s
and spirals but were chosen to be kept
as a part of sample of S0s, 
as descirbed in Section 2. The other remaining galaxy is NGC 4880
which is classified as $\rm{SA0}\wedge+(\rm{r})$. Its SDSS color composite image
suggests that
this galaxy may have had a spiral structure which is now nearly lost. All
these five objects can be thought of as having started as spirals which are now
in transition and may eventually acquire the morphology of a typical S0
galaxy.

The three spiral galaxies NGC 1341, NGC 4571 and NGC 4136 lying ``amidst'' S0s
in the $r_e - r_d$ diagram,
also share an interesting commonality. NGC 1341 is an SAab
galaxy residing in
the Fornax cluster, NGC 4571 is an SAd galaxy in the Virgo cluster and NGC
4136 is an SAc galaxy, known to be a member of a group. Apart from being in
rich environments, these three galaxies are observed to have anaemic spiral
arms and have been speculated to be transition objects between S0s and spirals
\citep{vandenBergh76, Kormendy2012}.

Another correlation of interest is that between the disk central brightness and
its scale length, also explored in \citet{Vaghmare2013}, who compared the
classical bulge hosts with pseudobulge
hosts and found that for a given disk scale length, pseudobulge hosts had a
lower central surface brightness. In Figure \ref{fig:mu0d_rd} we plot the same
diagram for the pseudobulge hosts of the two samples. As was the case with
Figure \ref{fig:rerd_pbs}, a trend is seen only along the disk scale length
axis while no trend is found on the disk central brightness axis. In other
words, the disk central brightness is the same on average for disks in both
populations. The difference is largely seen in scale lengths.

\section{Discussion}

We have presented what we believe is the first systematic comparison between
S0s and spiral galaxies hosting pseudobulges. In the previous section, we
have compared these two populations and found that
the disk scale length of S0s is lower on average than that of spirals, for a
given bulge effective radius. We further find that there is no significant
difference between these two populations with respect to disk central
brightness. A difference is seen in average surface brightness within
$r_e$. In this section, we discuss these results and try to explain them in the
context of various formation scenarios of both pseudobulges and S0 galaxies.

Before we discuss the results, we consider an important question
concerning the robustness of the results. While the samples have a similar
K-band luminosity (and hence mass) range and represent well the galaxies in the
nearby universe, the sample
of S0s has no cut-off on inclination while the sample of spirals does have such
a cut-off. One is thus
led to question whether this can explain the difference in scale lengths of
disks of pseudobulge hosts, which is the major result of the current study. We
could not find literature aimed at deriving inclination corrections to scale
lengths at 3.6 micron wavelength. So, we used the corrections for K-band used
by \citet{Laurikainen2010} which in turn are based on studies by
\citet{GrahamWorley2008} and \citet{Driver2008} and have ensured that the
differences in scalelengths cannot be explained as projection a effect as the
corrections are too small relative to the difference found. The mean correction
in the disk scale length is $\sim 0.02$ dex and as can be discerned from the
Figure \ref{fig:rerd_pbs}, is not sufficient enough to account for the observed
differences in the disk scale lengths. The differences in
corrections for 3.6 micron and K-band (2.2 micron) are expected to be
fairly small. We note that the parameters listed in the tables and those
used for plotting are not corrected for inclination effects. Even
bulge parameters are subject to projection effects, however determing
these is difficult. An example of such a study is the one carried out by
\citet{Pastrav2013}.

\subsection{Pseudobulge Fraction in Spiral Galaxies}
In the present study, we have used a sample of 31 spiral galaxies and using
both the S\'{e}rsic index and the position on the Kormendy diagram as 
classification crtieria, we classify 24 of these as pseudobulge hosts. This
fraction (77\%) is consistent with that found by \citet{Weinzirl2009}, for
example, who use optical imaging while we use mid-infrared imaging.
\citet{Lorenzo2014} used a sample of 189 isolated galaxies and found that
that 94\% of the bulges in their sample are
pseudobulges. A major reason for such a high fraction is due to their adopted
criterion - they classify all bulges with $n<2.5$ as pseudobulges, while we use
a more conservative classifcation scheme.
We have discussed in Section 3.1 the possible misclassification of
bulges and pseudobulges if the value of $n$ is used as the sole criterion.
\citet{Fisher2011} find that classical bulges only account for 12\% of the
total population of disk galaxies with the rest being pseudobulges or galaxies
with very low B/T
ratios. This high
fraction in turn implies that a large number of spiral galaxies have likely not
experienced evolution driven by external influences and that their evolution is
largely secular, with possibly some aid provided by minor mergers.
As pointed out by studies such as the one by \citet{Fisher2011}, this poses a
challenge to the currently accepted paradigm of galaxy formation where most
galaxies form through hierarchical clustering, through series of merger events.

\subsection{Lowered Scale Length - Role of Gas Stripping}

A common explanation invoked to explain the formation of S0 galaxies is that
they form as a result of spirals losing their spiral arms due to gas stripping.
The density-morphology relation
\citep{Dressler1980} has often been interpreted as evidence for this. However, 
\citet{Burstein2005} carefully measured total K-band absolute magnitudes of a
sample containing both S0s and spirals (both early- and late-type) and found
that S0s are more but not less luminous than spirals. This, according to the
authors,
rules out gas stripping as a viable mechanism as a difference of $\sim 0.75$
magnitudes, with S0s being fainter, is expected. We computed the difference in
the mean total magnitude for two populations of pseudobulge hosts and found that
pseudobulge hosting S0s
are indeed brighter by $\sim 0.24$ magnitudes. But if we divide our
pseudobulge hosting spirals into early-type ($T<5$, 12 galaxies) and
late-type($T \ge 5$, 19
galaxies), we find that pseudobulge hosting S0s are fainter by $\sim 1$
magnitude than pseudobulge hosting early-type
spirals while brighter by $\sim 1$ magnitude than pseudobulge hosting late-type
spirals. Thus
pseudobulge hosting early-type spirals transforming to pseudobulge hosting S0s
via gas stripping cannot be ruled out based on the arguments of
\citet{Burstein2005}. (A caveat in
this calculation is of course the overall low number of objects involved.)

Thus we see that we should not treat spiral galaxies as a single
population; early-type and late-type spirals have different properties and the
PB hosting early type spirals may in fact transform into PB hosting S0s via gas
stripping. It is also the case that we should not treat all the S0s as a single
population; they have different properties depending on luminosities as shown by
\citet{Barway2007, Barway2009} and on the bulge-type found in them, as shown by
\citet{Vaghmare2013}.

The disk scale length is a free parameter of an exponential function
which is used for describing the light distribution in a disk. The total flux
received from a disk with central surface brightness $\mu_{0d}$ and a disk scale
length $r_d$ is given by $\sim 2 \pi I_d(0) r_d^2$, where $I_d(0)$ is the
central intensity of the disk corresponding to $\mu_{0d}$. Thus a lowered scale
length as in the case of S0s would imply a lowered disk luminosity if $I_d(0)$
remains the same. This relation between disk luminosity and scale length can be
checked easily using Figure \ref{fig:dabsmag_rd}, which shows a plot of the disk
absolute magnitude against disk scale length. Unlike previous plots involving
disk scale length where trends were found only along the disk scale length axis,
this plot shows trends along the disk absolute magnitude as well. Disks of
pseudobulge hosting S0s tend to be less luminous than the disks in pseudobulge
hosting spirals. The mean disk absolute magnitudes for pseudobulge hosting S0s
and pseudobulge hosting spirals are $\sim -18.16$ and $-19.57$ respectively.
This difference of $1.41$ mangitudes is significant at a level better than
99.9\%. This is expected in a scenario where S0s have spirals as their
progenitors which undergo gas stripping to acquire a morphology resembling S0s.

We have shown that early-type pseudobulge hosting spirals cannot be ruled out as progenitors of
pseudobulge hosting S0s in the gas stripping scenario. We have also compared
disk absolute magnitudes to lend further support. Let us now consider the bulge
absolute magnitudes to try and answer the question whether gas stripping alone
is sufficient enough to bring about the transformation from early-type
pseudobulge hosting spirals to S0s. If gas stripping is the only process at
work, then one should not, for example, notice an increase in bulge luminosity.
The mean bulge absolute magnitudes for pseudobulge hosts in S0s and spirals are
respectively -17.16 and -16.45. If we treat the spirals separately as early- and
late-type, the mean values are -18.19 and -15.41 respectively. So, pseudobulges
of S0s are fainter than those of early-type spirals but brighter than those of
late-type spirals.

If early-type spirals hosting pseudobulges undergo gas stripping,
subsequent growth in the pseudobulge is also hindered - the lower luminosity of
pseudobulges in S0s compared to those in early-type, is consistent with this.
So, no additional process seems to be necessary to bring about this
transformation. In the case of pseudobulge hosting late-type spirals though,
this is not the case i.e. the S0 pseudobulges are more luminous. It is thus not
possible for gas stripping alone to transform pseudobulge hosting late-type
spirals into S0s unless an external process, such as a minor merger or an
accretion of a nearby dwarf builds up the bulge luminosity. In general, it seems
more viable for early-type pseudobulge hosting spirals to be the progenitors of
pseudobulge hosting S0s.

\subsection{Dynamical Formation Scenarios} By dynamical formation scenarios, we
refer to any mechanisms for forming pseudobulge hosting galaxies which are other
than (a) internal secular evolution and (b) evolution caused due to
environmental effects such as gas stripping. One such formation scenario, as
described in \citet{Guedes2013} and \citet{Okamoto2013}, is that pseudobulges
formed as inner disks at $z \sim 2$ through a series of star bursts and evolved
to their present day forms. However, it is difficult to verify whether such a
formation mechanism is viable for pseudobulges in our sample mainly because
there are no specific signatures predicted by these simulations. If any such
simulations predict, say, a low disk luminosity, one may be able to comment on
whether pseudobulges have formed in this manner. 

Another possible dynamical evolution scenario is that a pseudobulge of lower
than typical observed mass forms via secular processes and its eventual bulge
growth is caused through minor mergers, as demonstrated in \citet{Eliche2012},
who suggest that dry minor mergers can cause growth in the overall pseudobulge
mass without really affecting scaling relations such as the $r_e - r_d$
correlation, provided such a correlation already exists. This opens the
possibility for late type spirals to evolve into S0s too but that will still
require some process to enable the removal of the spiral arms. 

\section{Summary} We have presented a comparison of pseudobulges in S0 and
spiral galaxies using structural parameters derived from 2-d decomposition of
mid-infrared images taken at 3.6$\mu$m by Spitzer IRAC. We find that among
spiral galaxies, 77\% of the bulges are classified as pseudobulges. As pointed
out by various studies, the presence of such a large fraction poses problems to
our current picture of galaxy formation. However, our primary result is that the
disk scale length of pseudobulge hosting S0s is significantly smaller on average
than that of their spiral counterparts. This can be explained as a lowered disk
luminosity which in turn implies that S0s have evolved from spiral progenitors.
We also argue that early type spirals are more likely to be the progenitors
based on total luminosity arguments. We speculate that if late type spirals
hosting pseudobulges have to evolve into S0s, a mechanism other than gas
stripping of spirals is needed.

We have also investigated the effect of environment on pseudobulges in
the two samples. But we did not find any significant trends in the properties of
the pseudobulges as a function of the various structural parameters. The study
is made more difficult because of the low number statistics one deals with when
the sample is sub-divided based on whether it is in a field or group/cluster
environment. We are currently in the process of planning a study to investigate
pseudobulge hosting properties as a function of the environment using a sample
containing pseudobulges occuring in a variety of environments.

\section*{ACKNOWLEDGEMENTS} We would like to thank the anonymous referee whose
comments helped improve the content and presentation of the manuscript.  KV
acknowledges Council of Scientific and Industrial Research (CSIR), India for
financial support. SB would like to acknowledge a bilateral grant under the
Indo-South Africa Science and Technology Cooperation (UID-76354) funded by
Departments of Science and Technology (DST) of the Indian and South African
Governments. This paper is based upon work supported financially by the National
Research Foundation (NRF), South Africa. Any opinions, findings and conclusions
or recommendations expressed in this paper are those of the authors and
therefore the NRF does not accept any liability in regard thereto. SM
acknowledges hospitality at IUCAA during her visit, where initial plans of this
project were made. We thank Yogesh Wadadekar and Kanak Saha for discussions.

This work is based in part on observations made with the Spitzer Space
Telescope, which is operated by the Jet Propulsion Laboratory, California
Institute of Technology under a contract with NASA. This research has made use
of the NASA/IPAC Extragalactic Database (NED) which is operated by the Jet
Propulsion Laboratory, California Institute of Technology, under contract with
the National Aeronautics and Space Administration. We acknowledge the usage of
the HyperLeda database (http://leda.univ-lyon1.fr). The research also makes
extensive use of iPython, an interactive Python based scientific computing
environment \citep{Perez} and Astropy, a community-developed core Python package
for Astronomy \citet{astropy}.

\bibliographystyle{mn2e} \bibliography{Bibliography_Master.bib} 
\begin{thebibliography}{58}
\expandafter\ifx\csname natexlab\endcsname\relax\def\natexlab#1{#1}\fi

\bibitem[{{Aguerri} {et~al.}(2005){Aguerri}, {Iglesias-P{\'a}ramo},
  {V{\'{\i}}lchez}, {Mu{\~n}oz-Tu{\~n}{\'o}n}, \&
  {S{\'a}nchez-Janssen}}]{Aguerri2005}
{Aguerri} J.~A.~L., {Iglesias-P{\'a}ramo} J., {V{\'{\i}}lchez} J.~M.,
  {Mu{\~n}oz-Tu{\~n}{\'o}n} C., {S{\'a}nchez-Janssen} R., 2005, \aj, 130, 475

\bibitem[{{Arag{\'o}n-Salamanca} {et~al.}(2006){Arag{\'o}n-Salamanca},
  {Bedregal}, \& {Merrifield}}]{Salamanca2006}
{Arag{\'o}n-Salamanca} A., {Bedregal} A.~G., {Merrifield} M.~R., 2006, \aap,
  458, 101

\bibitem[{{Astropy Collaboration} {et~al.}(2013){Astropy Collaboration},
  {Robitaille}, {Tollerud}, {Greenfield}, {Droettboom}, {Bray}, {Aldcroft},
  {Davis}, {Ginsburg}, {Price-Whelan}, {Kerzendorf}, {Conley}, {Crighton},
  {Barbary}, {Muna}, {Ferguson}, {Grollier}, {Parikh}, {Nair}, {Unther},
  {Deil}, {Woillez}, {Conseil}, {Kramer}, {Turner}, {Singer}, {Fox}, {Weaver},
  {Zabalza}, {Edwards}, {Azalee Bostroem}, {Burke}, {Casey}, {Crawford},
  {Dencheva}, {Ely}, {Jenness}, {Labrie}, {Lim}, {Pierfederici}, {Pontzen},
  {Ptak}, {Refsdal}, {Servillat}, \& {Streicher}}]{astropy}
{Astropy Collaboration}, {Robitaille} T.~P., {Tollerud} E.~J., {Greenfield} P.,
  {Droettboom} M., {Bray} E., {Aldcroft} T., {Davis} M., {Ginsburg} A.,
  {Price-Whelan} A.~M., {Kerzendorf} W.~E., {Conley} A., {Crighton} N.,
  {Barbary} K., {Muna} D., {Ferguson} H., {Grollier} F., {Parikh} M.~M., {Nair}
  P.~H., {Unther} H.~M., {Deil} C., {Woillez} J., {Conseil} S., {Kramer} R.,
  {Turner} J.~E.~H., {Singer} L., {Fox} R., {Weaver} B.~A., {Zabalza} V.,
  {Edwards} Z.~I., {Azalee Bostroem} K., {Burke} D.~J., {Casey} A.~R.,
  {Crawford} S.~M., {Dencheva} N., {Ely} J., {Jenness} T., {Labrie} K., {Lim}
  P.~L., {Pierfederici} F., {Pontzen} A., {Ptak} A., {Refsdal} B., {Servillat}
  M., {Streicher} O., 2013, \aap, 558, A33

\bibitem[{{Athanassoula}(1992)}]{Athanasoula1992}
{Athanassoula} E., 1992, \mnras, 259, 345

\bibitem[{{Barr} {et~al.}(2007){Barr}, {Bedregal}, {Arag{\'o}n-Salamanca},
  {Merrifield}, \& {Bamford}}]{Barr2007}
{Barr} J.~M., {Bedregal} A.~G., {Arag{\'o}n-Salamanca} A., {Merrifield} M.~R.,
  {Bamford} S.~P., 2007, \aap, 470, 173

\bibitem[{{Barway} {et~al.}(2007){Barway}, {Kembhavi}, {Wadadekar},
  {Ravikumar}, \& {Mayya}}]{Barway2007}
{Barway} S., {Kembhavi} A., {Wadadekar} Y., {Ravikumar} C.~D., {Mayya} Y.~D.,
  2007, \apjl, 661, L37

\bibitem[{{Barway} {et~al.}(2009){Barway}, {Wadadekar}, {Kembhavi}, \&
  {Mayya}}]{Barway2009}
{Barway} S., {Wadadekar} Y., {Kembhavi} A.~K., {Mayya} Y.~D., 2009, \mnras,
  394, 1991

\bibitem[{{Bedregal} {et~al.}(2006){Bedregal}, {Arag{\'o}n-Salamanca}, \&
  {Merrifield}}]{Bedregal2006}
{Bedregal} A.~G., {Arag{\'o}n-Salamanca} A., {Merrifield} M.~R., 2006, \mnras,
  373, 1125

\bibitem[{{Bertin} \& {Arnouts}(1996)}]{Bertin1996}
{Bertin} E., {Arnouts} S., 1996, \aaps, 117, 393

\bibitem[{{Burstein} {et~al.}(2005){Burstein}, {Ho}, {Huchra}, \&
  {Macri}}]{Burstein2005}
{Burstein} D., {Ho} L.~C., {Huchra} J.~P., {Macri} L.~M., 2005, \apj, 621, 246

\bibitem[{{Capaccioli}(1989)}]{Capaccioli89}
{Capaccioli} M., 1989, in World of Galaxies (Le Monde des Galaxies), {Corwin}
  Jr. H.~G., {Bottinelli} L., eds., pp. 208--227

\bibitem[{{Carollo} {et~al.}(2001){Carollo}, {Stiavelli}, {de Zeeuw}, {Seigar},
  \& {Dejonghe}}]{Carollo2001}
{Carollo} C.~M., {Stiavelli} M., {de Zeeuw} P.~T., {Seigar} M., {Dejonghe} H.,
  2001, \apj, 546, 216

\bibitem[{{Carollo} {et~al.}(1998){Carollo}, {Stiavelli}, \&
  {Mack}}]{Carollo98}
{Carollo} C.~M., {Stiavelli} M., {Mack} J., 1998, \aj, 116, 68

\bibitem[{{Courteau} {et~al.}(1996){Courteau}, {de Jong}, \&
  {Broeils}}]{Courteau96}
{Courteau} S., {de Jong} R.~S., {Broeils} A.~H., 1996, \apjl, 457, L73

\bibitem[{{de Vaucouleurs} {et~al.}(1991){de Vaucouleurs}, {de Vaucouleurs},
  {Corwin}, {Buta}, {Paturel}, \& {Fouqu{\'e}}}]{deVauc91}
{de Vaucouleurs} G., {de Vaucouleurs} A., {Corwin} Jr. H.~G., {Buta} R.~J.,
  {Paturel} G., {Fouqu{\'e}} P., 1991, {Third Reference Catalogue of Bright
  Galaxies. Volume I: Explanations and references. Volume II: Data for galaxies
  between 0$^{h}$ and 12$^{h}$. Volume III: Data for galaxies between 12$^{h}$
  and 24$^{h}$.}

\bibitem[{{Debattista} {et~al.}(2006){Debattista}, {Mayer}, {Carollo}, {Moore},
  {Wadsley}, \& {Quinn}}]{Debattista2006}
{Debattista} V.~P., {Mayer} L., {Carollo} C.~M., {Moore} B., {Wadsley} J.,
  {Quinn} T., 2006, \apj, 645, 209

\bibitem[{{Djorgovski} \& {Davis}(1987)}]{Djorgovski87}
{Djorgovski} S., {Davis} M., 1987, \apj, 313, 59

\bibitem[{{Dressler}(1980)}]{Dressler1980}
{Dressler} A., 1980, \apj, 236, 351

\bibitem[{{Dressler} {et~al.}(1987){Dressler}, {Lynden-Bell}, {Burstein},
  {Davies}, {Faber}, {Terlevich}, \& {Wegner}}]{Dressler87}
{Dressler} A., {Lynden-Bell} D., {Burstein} D., {Davies} R.~L., {Faber} S.~M.,
  {Terlevich} R., {Wegner} G., 1987, Astrophysical Journal, 313, 42

\bibitem[{{Driver} {et~al.}(2008){Driver}, {Popescu}, {Tuffs}, {Graham},
  {Liske}, \& {Baldry}}]{Driver2008}
{Driver} S.~P., {Popescu} C.~C., {Tuffs} R.~J., {Graham} A.~W., {Liske} J.,
  {Baldry} I., 2008, \apjl, 678, L101

\bibitem[{{Drory} \& {Fisher}(2007)}]{Drory2007}
{Drory} N., {Fisher} D.~B., 2007, \apj, 664, 640

\bibitem[{{Eliche-Moral} {et~al.}(2012){Eliche-Moral},
  {Gonz{\'a}lez-Garc{\'{\i}}a}, {Aguerri}, {Gallego}, {Zamorano}, {Balcells},
  \& {Prieto}}]{Eliche2012}
{Eliche-Moral} M.~C., {Gonz{\'a}lez-Garc{\'{\i}}a} A.~C., {Aguerri} J.~A.~L.,
  {Gallego} J., {Zamorano} J., {Balcells} M., {Prieto} M., 2012, \aap, 547, A48

\bibitem[{{Elmegreen} {et~al.}(2009){Elmegreen}, {Elmegreen}, {Fernandez}, \&
  {Lemonias}}]{Elmegreen2009}
{Elmegreen} B.~G., {Elmegreen} D.~M., {Fernandez} M.~X., {Lemonias} J.~J.,
  2009, \apj, 692, 12

\bibitem[{{Falc{\'o}n-Barroso} {et~al.}(2011){Falc{\'o}n-Barroso}, {van de
  Ven}, {Peletier}, {Bureau}, {Jeong}, {Bacon}, {Cappellari}, {Davies}, {de
  Zeeuw}, {Emsellem}, {Krajnovi{\'c}}, {Kuntschner}, {McDermid}, {Sarzi},
  {Shapiro}, {van den Bosch}, {van der Wolk}, {Weijmans}, \& {Yi}}]{Falcon2011}
{Falc{\'o}n-Barroso} J., {van de Ven} G., {Peletier} R.~F., {Bureau} M.,
  {Jeong} H., {Bacon} R., {Cappellari} M., {Davies} R.~L., {de Zeeuw} P.~T.,
  {Emsellem} E., {Krajnovi{\'c}} D., {Kuntschner} H., {McDermid} R.~M., {Sarzi}
  M., {Shapiro} K.~L., {van den Bosch} R.~C.~E., {van der Wolk} G., {Weijmans}
  A., {Yi} S., 2011, \mnras, 417, 1787

\bibitem[{{Fisher} \& {Drory}(2008)}]{Fisher2008}
{Fisher} D.~B., {Drory} N., 2008, \aj, 136, 773

\bibitem[{{Fisher} \& {Drory}(2010)}]{Fisher2010}
---, 2010, \apj, 716, 942

\bibitem[{{Fisher} \& {Drory}(2011)}]{Fisher2011}
---, 2011, \apjl, 733, L47

\bibitem[{{Gadotti}(2008)}]{Gadotti2008}
{Gadotti} D.~A., 2008, \mnras, 384, 420

\bibitem[{{Gadotti}(2009)}]{Gadotti2009}
---, 2009, \mnras, 393, 1531

\bibitem[{{Graham} \& {Worley}(2008)}]{GrahamWorley2008}
{Graham} A.~W., {Worley} C.~C., 2008, \mnras, 388, 1708

\bibitem[{{Guedes} {et~al.}(2013){Guedes}, {Mayer}, {Carollo}, \&
  {Madau}}]{Guedes2013}
{Guedes} J., {Mayer} L., {Carollo} M., {Madau} P., 2013, \apj, 772, 36

\bibitem[{{Hubble}(1936)}]{Hubble1936}
{Hubble} E.~P., 1936, {Realm of the Nebulae}

\bibitem[{{Immeli} {et~al.}(2004){Immeli}, {Samland}, {Gerhard}, \&
  {Westera}}]{Immeli2004}
{Immeli} A., {Samland} M., {Gerhard} O., {Westera} P., 2004, \aap, 413, 547

\bibitem[{{Keselman} \& {Nusser}(2012)}]{Keselman2012}
{Keselman} J.~A., {Nusser} A., 2012, \mnras, 424, 1232

\bibitem[{{Khosroshahi} {et~al.}(2000){Khosroshahi}, {Wadadekar}, \&
  {Kembhavi}}]{Khosroshahi2000}
{Khosroshahi} H.~G., {Wadadekar} Y., {Kembhavi} A., 2000, \apj, 533, 162

\bibitem[{{Kormendy}(1977)}]{Kormendy77}
{Kormendy} J., 1977, \apj, 218, 333

\bibitem[{{Kormendy} \& {Bender}(2012)}]{Kormendy2012}
{Kormendy} J., {Bender} R., 2012, \apjs, 198, 2

\bibitem[{{Kormendy} \& {Kennicutt}(2004)}]{Kormendy2004}
{Kormendy} J., {Kennicutt} Jr. R.~C., 2004, \araa, 42, 603

\bibitem[{{Laurikainen} {et~al.}(2005){Laurikainen}, {Salo}, \&
  {Buta}}]{Laurikainen2005}
{Laurikainen} E., {Salo} H., {Buta} R., 2005, \mnras, 362, 1319

\bibitem[{{Laurikainen} {et~al.}(2010){Laurikainen}, {Salo}, {Buta}, {Knapen},
  \& {Comer{\'o}n}}]{Laurikainen2010}
{Laurikainen} E., {Salo} H., {Buta} R., {Knapen} J.~H., {Comer{\'o}n} S., 2010,
  \mnras, 405, 1089

\bibitem[{Lorenzo {et~al.}(2014)Lorenzo, Sulentic, Verdes-Montenegro,
  Blasco-Herrera, Argudo-Fernández, Garrido, Ramírez-Moreta, Ruiz,
  Sánchez-Expósito, \& Santander-Vela}]{Lorenzo2014}
Lorenzo M.~F., Sulentic J., Verdes-Montenegro L., Blasco-Herrera J.,
  Argudo-Fernández M., Garrido J., Ramírez-Moreta P., Ruiz J.~E.,
  Sánchez-Expósito S., Santander-Vela J.~D., 2014, The Astrophysical Journal
  Letters, 788, L39

\bibitem[{{Mathur} {et~al.}(2012){Mathur}, {Fields}, {Peterson}, \&
  {Grupe}}]{Mathur2012}
{Mathur} S., {Fields} D., {Peterson} B.~M., {Grupe} D., 2012, \apj, 754, 146

\bibitem[{{Mu{\~n}oz-Mateos} {et~al.}(2009){Mu{\~n}oz-Mateos}, {Gil de Paz},
  {Zamorano}, {Boissier}, {Dale}, {P{\'e}rez-Gonz{\'a}lez}, {Gallego},
  {Madore}, {Bendo}, {Boselli}, {Buat}, {Calzetti}, {Moustakas}, \&
  {Kennicutt}}]{Munoz2009}
{Mu{\~n}oz-Mateos} J.~C., {Gil de Paz} A., {Zamorano} J., {Boissier} S., {Dale}
  D.~A., {P{\'e}rez-Gonz{\'a}lez} P.~G., {Gallego} J., {Madore} B.~F., {Bendo}
  G., {Boselli} A., {Buat} V., {Calzetti} D., {Moustakas} J., {Kennicutt} Jr.
  R.~C., 2009, \apj, 703, 1569

\bibitem[{{Noguchi}(1999)}]{Noguchi99}
{Noguchi} M., 1999, \apj, 514, 77

\bibitem[{{Okamoto}(2013)}]{Okamoto2013}
{Okamoto} T., 2013, \mnras, 428, 718

\bibitem[{{Oke} \& {Gunn}(1983)}]{OkeGunn83}
{Oke} J.~B., {Gunn} J.~E., 1983, \apj, 266, 713

\bibitem[{{Orban de Xivry} {et~al.}(2011){Orban de Xivry}, {Davies},
  {Schartmann}, {Komossa}, {Marconi}, {Hicks}, {Engel}, \&
  {Tacconi}}]{Xivry2011}
{Orban de Xivry} G., {Davies} R., {Schartmann} M., {Komossa} S., {Marconi} A.,
  {Hicks} E., {Engel} H., {Tacconi} L., 2011, \mnras, 417, 2721

\bibitem[{{Pastrav} {et~al.}(2013){Pastrav}, {Popescu}, {Tuffs}, \&
  {Sansom}}]{Pastrav2013}
{Pastrav} B.~A., {Popescu} C.~C., {Tuffs} R.~J., {Sansom} A.~E., 2013, \aap,
  557, A137

\bibitem[{{Peng} {et~al.}(2002){Peng}, {Ho}, {Impey}, \& {Rix}}]{Peng2002}
{Peng} C.~Y., {Ho} L.~C., {Impey} C.~D., {Rix} H.-W., 2002, \aj, 124, 266

\bibitem[{P\'erez \& Granger(2007)}]{Perez}
P\'erez F., Granger B.~E., 2007, Computing in Science and Engineering, 9, 21

\bibitem[{{Ravikumar} {et~al.}(2006){Ravikumar}, {Barway}, {Kembhavi},
  {Mobasher}, \& {Kuriakose}}]{Ravikumar2006}
{Ravikumar} C.~D., {Barway} S., {Kembhavi} A., {Mobasher} B., {Kuriakose}
  V.~C., 2006, \aap, 446, 827

\bibitem[{{Renzini}(1999)}]{Renzini99}
{Renzini} A., 1999, in The Formation of Galactic Bulges, {Carollo} C.~M.,
  {Ferguson} H.~C., {Wyse} R.~F.~G., eds., p.~9

\bibitem[{{Sersic}(1968)}]{Sersic68}
{Sersic} J.~L., 1968, {Atlas de galaxias australes}

\bibitem[{{Toloba} {et~al.}(2012){Toloba}, {Boselli}, {Peletier},
  {Falc{\'o}n-Barroso}, {van de Ven}, \& {Gorgas}}]{Toloba2012}
{Toloba} E., {Boselli} A., {Peletier} R.~F., {Falc{\'o}n-Barroso} J., {van de
  Ven} G., {Gorgas} J., 2012, \aap, 548, A78

\bibitem[{{Tully} \& {Fisher}(1988)}]{Tully1988}
{Tully} R.~B., {Fisher} J.~R., 1988, {Catalog of Nearby Galaxies}

\bibitem[{{Vaghmare} {et~al.}(2013){Vaghmare}, {Barway}, \&
  {Kembhavi}}]{Vaghmare2013}
{Vaghmare} K., {Barway} S., {Kembhavi} A., 2013, \apjl, 767, L33

\bibitem[{{van den Bergh}(1976)}]{vandenBergh76}
{van den Bergh} S., 1976, \apj, 206, 883

\bibitem[{{Weinzirl} {et~al.}(2009){Weinzirl}, {Jogee}, {Khochfar}, {Burkert},
  \& {Kormendy}}]{Weinzirl2009}
{Weinzirl} T., {Jogee} S., {Khochfar} S., {Burkert} A., {Kormendy} J., 2009,
  \apj, 696, 411

\end{thebibliography}
\end{document}